# Chirality of plasmonic metasurfaces with rectangular holes

Biyuan Wu[1,2], Mingjun Wang[1], Yasong Sun[3,4,*], Feng Wu[5], Zhangxing Shi[2], and

Xiaohu Wu[2,*]

[1]*School of Automation and Information Engineering, Xi'an University of Technology, Shaanxi,*

*Xi'an 710048, China*

[2]*Shandong Institute of Advanced Technology, Jinan 250100, China*

[3]*Basic Research Center, School of Power and Energy, Northwestern Polytechnical University,*

*Shaanxi, Xi'an 710064, China*

[4]*Center of Computational Physics and Energy Science, Yangtze River Delta Research Institute of*

*NPU, Northwestern Polytechnical University, Jiangsu, Taicang 215400, China*

[5]*School of Optoelectronic Engineering, Guangdong Polytechnic Normal University, Guangzhou*

*510665, China*

*\*Corresponding authors: yssun@nwpu.edu.cn (Y. Sun);*

*xiaohu.wu@iat.cn (X. Wu)*

# Abstract

Chiral response is of tremendous importance to many fields, such as analytical chemistry, polarization manipulation and biological sensing. Here, a chiral metasurface based on rectangular holes is systematically investigated. The results show that the chirality is closely related to the size and the orientation of resonance unit. It is found that the period of the structure is always smaller than the wavelength at which chirality appears, which will provide a good basis for the design of chiral structures. More importantly, the CD is highly sensitive to the orientation of resonance unit. By adjusting the rotation angle, it is not only possible to invert the CD, but also to change the symmetry of the structure to realize the regulation of chirality. The chirality can be significantly enhanced in the proposed structure, and the maximum of circular dichroism (CD) can reach 0.76. To better understand the physical mechanism, the distributions of electric field for LCP and RCP waves are also discussed as well. This work will not only deepen the understanding of chiral metasurfaces, but also provide guidance for the design of similar chiral structures.

## 1. Introduction

Light-matter interactions and manipulation of light polarization states has long been a hot research area in physics [1]. Circularly polarized wave, a fundamental polarization state, is often considered to be a macroscopic manifestation of a particular spin photon set. It turns out that manipulating the spin of electrons with circularly polarized light plays a critical role in quantum information technology [2]. Therefore, effective manipulation and detection of circularly polarized light is very important in many fields, such as quantum optics [3], chemistry [4] and biology [5].

Chiral objects cannot be overlapped with their mirror image by translation, scaling or rotation operations, and response differently to left-handed circularly polarized (LCP) wave and right-handed circularly polarized (RCP) wave [6-8]. The difference in reflection/transmission for LCP wave and RCP wave refers to circular dichroism (CD), which is an important parameter to characterize chirality. In nature, there are many chiral structures, such as protein molecules, helical structures, quartz crystal molecular structures, etc. [9]. However, the circular dichroism of natural chiral materials is very weak, which seriously restricts the practical applications [10]. Therefore, it is necessary to involve artificial dielectric and plasmonic structures to enhance circular dichroism.

To strongly enhance the chiral response, many chiral structures have been proposed [11-15]. Especially, chiral metamaterials have attracted enormous attention [16-18]. In terms of structural composition, chiral metamaterials can be divided into three-dimensional (3D) metamaterials [19-24] and two-dimensional (2D) planar structures [25-28]. In general, greater CD can be triggered in 3D nanostructures due to the strong near-field electromagnetic coupling between their multiple layers. For example, Wang et al. proposed bilayer metamaterials with a relative rotation angle to simultaneously break the rotational and mirror symmetries, leading to strong chirality [29]. It can be seen that the chirality in the nanostructures can be excited by introducing some asymmetry in the fabrication process or by appropriate settings. However, 3D chiral structures encounter great challenges in the fabrication procedures of their complex morphologies. Chiral metasurfaces with a single patterned layer have been extensively investigated as S-shaped [30] and L-shaped structures [14] since they can also exhibit strong chiral response and are relatively easier to fabricate compared with 3D nonplanar metamaterials. Recently, Ouyang et al. proposed a chiral metasurface absorber made of multiple double-rectangle resonators with different sizes to realize large CD [26]. Although many chiral metasurface structures have been proposed, their design principles are not clear enough and deserve further research.

In this paper, a chiral metasurface is proposed and demonstrated to achieve strong CD with 0.76. The metasurface is designed with metal-dielectric-metal structure including the top Au layer with rectangular holes, the middle $SiO_2$ dielectric layer, and the Au mirror layer. Firstly, we discuss the effect of structure period on the CD. It is found that the chirality can only be obtained when the period is smaller than the wavelength. Further, a strong chiral response is obtained by optimizing structural parameters of the rectangular holes, such as rotation angle, length and width. The electric field distributions under circular polarizations are simulated to explain the mechanism of the strong CD. More importantly, it is found that the CD of the proposed structure is zero with the rotation angle equal to 0 ° or an integral multiple of 30°. Finally, the arrangement of rectangular holes is replaced as square, which further verifies our conclusions. The demonstrated chiral metasurfaces based on rectangular holes possesses a strong ability to distinguish between LCP and RCP waves and will be very promising for practical applications in polarization manipulation, optical communication of spin information, and biological sensing.

## 2. Modeling and methods

Figure 1 illustrates schematic diagrams of the proposed chiral structure, which is composed of three layers. The top layer is Au film with periodic rectangular holes. The period of the rectangular holes is P. The thickness of the top layer is $h_1$. As shown in Fig. 1(b), all rectangular holes have width $w$ and length $t$, and the rotation angle with respect to the $x$ axis is $\theta$. The middle layer is SiO$_2$ with the thickness $h_2$. The refractive index of SiO$_2$ is 1.45 [31]. The bottom layer is Au film, whose thickness $h_3$ is larger than the skinning depth to prevent the transmitted wave. The permittivity of the Au is obtained from Johnson and Christy [32].

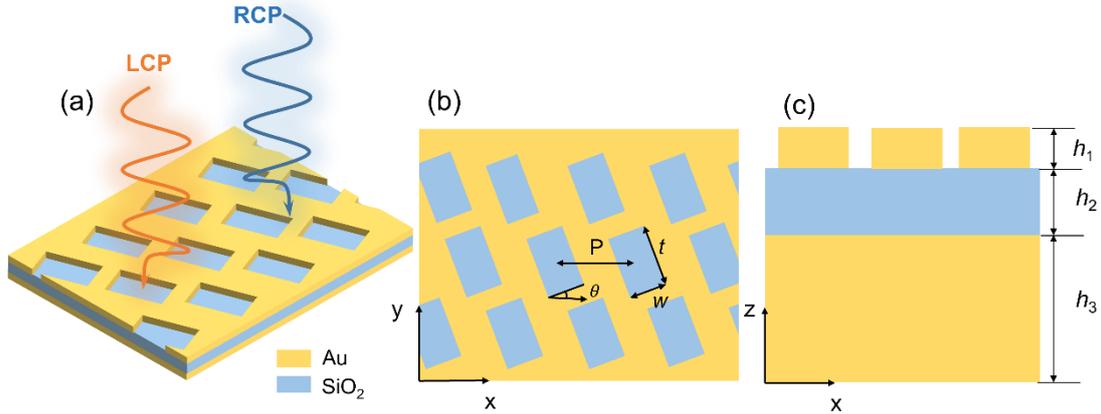

Fig. 1 Schematic diagrams of the proposed chiral metasurfaces. (a) 3D diagram; (b) top view; (c) side view.

The structure is illuminated by circularly polarized waves along $z$ direction. The reflection for LCP and RCP waves is represented by $R_{LCP}$ and $R_{RCP}$, respectively. Therefore, the circular dichroism can be calculated by:

$$CD = |R_{LCP} - R_{RCP}|. \qquad (1)$$

Finite-difference time-domain (FDTD) is employed to numerically calculate the reflection spectra for circularly polarized wave incidences. In the simulation, periodic boundary conditions are used in $x$ and $y$ directions and perfectly matching layers (PML) are applied in $z$ direction. Mesh type is auto non-uniform, and the mesh accuracy is set as 5. Circularly polarized wave is treated as the superposition of two linearly polarized waves of equal amplitude and 90° difference in phase. A monitor, i.e., frequency-domain field and power, is added to calculate the reflection.

## 3. Results and discussions

Firstly, the length and width of rectangular holes are fixed at 400 nm and 280 nm, respectively. As shown in Fig. 2, we discuss the influence of period on CD under different rotation angles. When the rotation angle is equal to 20°, it is found that the CD spectra occurs red shift and the maximum of CD gradually decreases as the period increases. Due to the diminishing near-field effect between the iholes as the period increases, which in turn leads to weak chirality. Besides, one can see that the wavelength is always larger than period of the structure when the chirality occurs at any rotation angles. For example, strong chirality occurs at wavelength 880 nm when the period is 500 nm. The period is smaller than the wavelength in this case. A similar phenomenon can be observed when the rotation angle is 40° and 80°. However, as can be clearly seen in Fig. 2(c), the CD is extremely small at any period when the period is 60°. This is a very peculiar phenomenon that deserves further exploration.

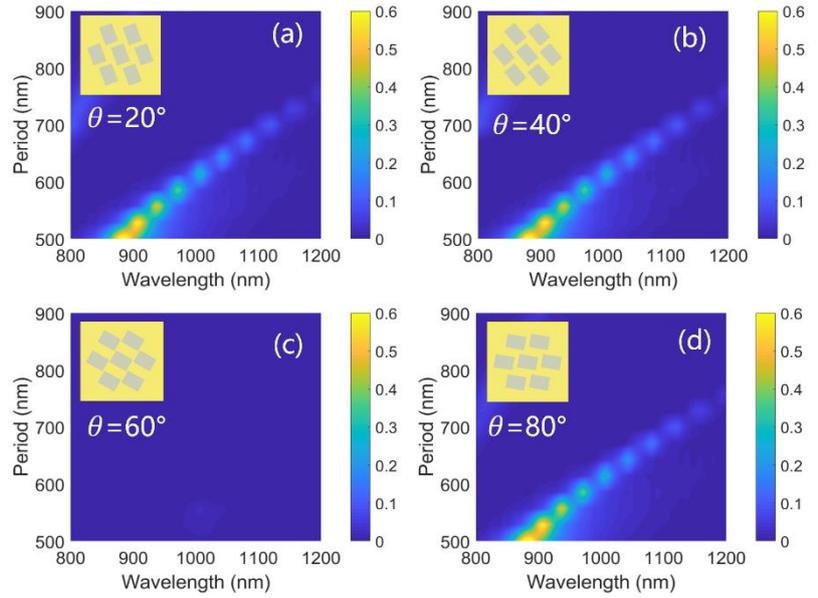

Fig. 2 The CD as functions of period and wavelength under different rotation angles: (a) 20°; (b) 40°; (c) 60°; (d) 80°. The length and width of rectangular holes are set to be $t$=400 nm and $w$=280 nm.

Now, the structure parameters are set as follows: P=500 nm, $\theta$=20°, $t$=400 nm and $w$=280 nm, respectively. Figure 3(a) shows the reflection of LCP and RCP waves and the corresponding CD. It is found that the structure can maintain high reflection from 800 nm to 1200 nm when RCP wave is incident while the reflection is only about 0.1 for LCP wave at wavelength 880 nm. Thus, there is a great difference in reflection between LCP wave and RCP wave at wavelength 880 nm, resulting in strong chirality. To understand the physical mechanism of selective reflection for LCP and RCP waves, shown in Fig. 3(b), the polarization conversion between LCP and RCP waves is calculated by CST studio suit 2019. Here the LCP-LCP and LCP-RCP represent the LCP wave component and RCP wave component of the reflected wave for LCP wave incidence, respectively. RCP-RCP and RCP-LCP have the similar definitions. It can clearly see that the reflection from LCP to LCP and RCP to RCP is identical. For an ordinary metallic mirror, there is only LCP (RCP) component in the reflected wave when LCP (RCP) wave is incident. For the proposed chiral structure, however, there are both LCP and RCP in the reflected wave when the circularly polarized wave with any handedness is incident. From Fig. 3(b), one can see that the reflected RCP wave is only 0.2 at resonance wavelength for LCP wave incidence, while the reflected RCP wave is greater than 0.8. Therefore, the chirality is attributed to the different polarization conversion between LCP and RCP waves.

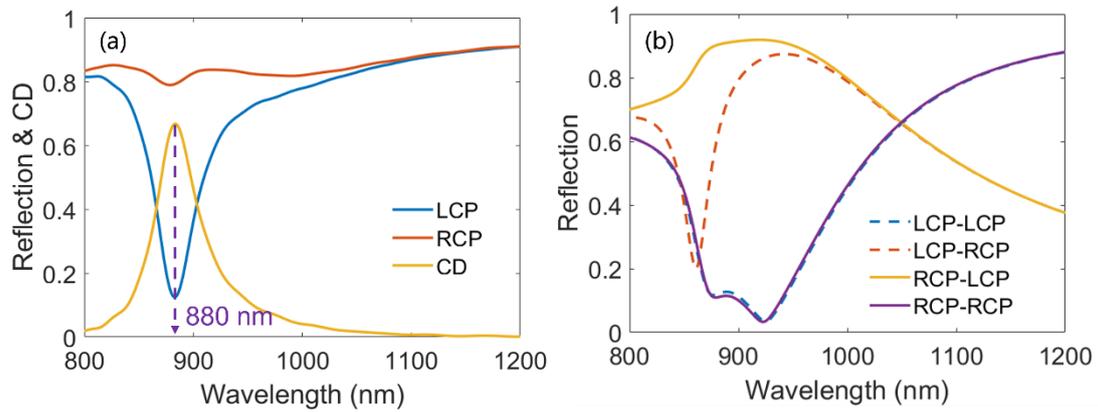

Fig. 3 (a) The reflection of LCP wave (blue line) and RCP wave (red line), and the corresponding CD (yellow line). (b) The polarization conversion between LCP and RCP waves. The structure parameters are set to be P=500 nm, $\theta$ =20°, t=400 nm and w=280 nm.

Based on the analysis of Fig. 2, it is found that the rotation angle is one of the key factors effecting chirality. To obtain stronger chirality, we investigate the effect of rotation angle on CD with a fixed incident wavelength 880 nm and period 500 nm. It can be found from Fig. 4(a) the selective reflection of the structure on the LCP wave and RCP wave can be reversed as the rotation angle increases from 0° to 90°, implying that the structure can turn the CD over via adjusting the rotation angle. Furthermore, it is found that the CD is equal to 0 at rotation angles 0°, 30°, 60° and 90°. However, three significant CDs appear at rotation angles 21°, 41°, and 81°, respectively. In order to study whether these three angles are special, we plot the angle-dependent CD curves as Fig. 4(b) with two different $t$ (330 and 350 nm) and a fixed $w$ (280 nm). Obviously, when $w$ is kept constant 280 nm and $t$ is equal to 330 nm or 350 nm, the rotation angle corresponding to the maximum value of CD is 13° or 17°, respectively. The angles of 21°, 41° and 81° mentioned above are therefore not special, but are closely related to the setting of other structural parameters. However, it can be seen that the CD still is 0 at different lengths of rectangular holes when the rotation angles are 0° and 30°.

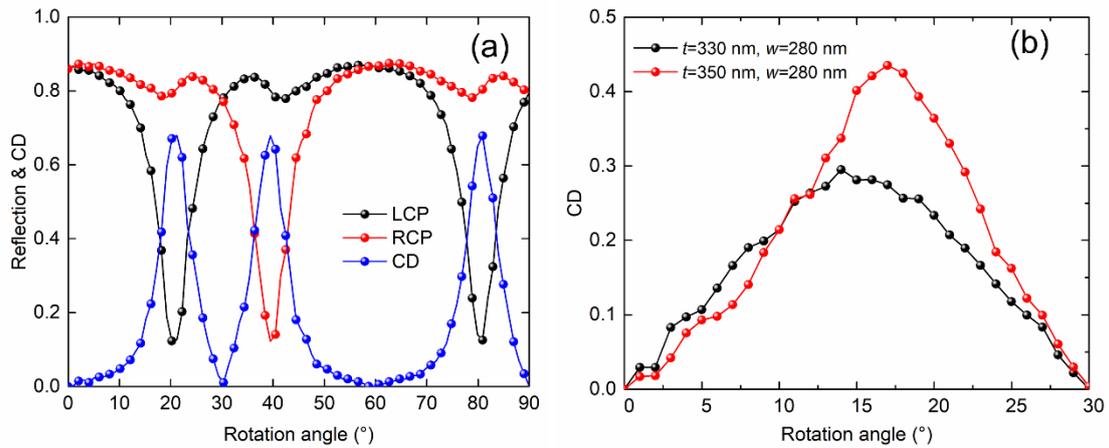

Fig. 4 (a) The reflection and CD as a function of rotation angle when P=500 nm, $t$=400 nm, $w$=280 nm, and the wavelength is 880 nm. (b) The CD as a function of rotation angle under different structural parameters.

The width $w$ and length $t$ of rectangular holes also play important roles in chirality. Here, the wavelength we studied is still 880 nm. When the rotation angle is 21°, Figs. 5(a) and 5(b) show the reflection as functions of $w$ and $t$ for LCP and RCP waves, respectively. One can see that the structure exhibits strong reflection when the width is smaller than 150 nm for both LCP and RCP waves. When $w$ and $t$ are greater than 250 nm and 350 nm, respectively, significant differences in the reflection of LCP and RCP wave appear. Thus, strong CD can be observed in Fig. 5(c), and the maximum of CD can reach 0.76 when $t$=400 nm and $w$=285.7 nm. Besides, we also discuss the effect of different sizes of rectangular holes on the CD when the rotation angle is 60°, shown in Figs. 5(d)-5(f). It clearly that the CD is almost equal to 0 at any sizes of rectangular holes, indicating that the chirality can be effectively regulated by rotating the rotation angle of the rectangle holes.

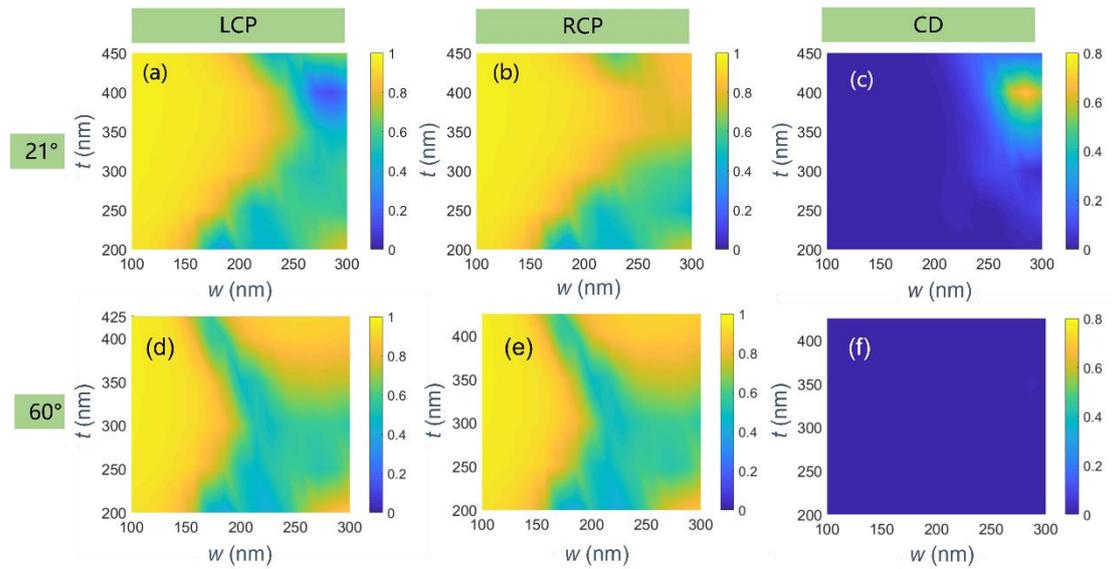

Fig. 5 The reflection as functions of $t$ and $w$ for (a, d) LCP wave and (b, e) RCP wave; (c, d) the corresponding CD. The wavelength is 880 nm and the rotation angle is (a, b, c) 21° and (d, e ,f) 60°, respectively.

To better understand the physical mechanism of the chiral response, the distributions of electric field in the $x$-$y$ plane at wavelength 880 nm are plotted. There are $t$=400 nm and $w$=285.7 nm .When rotation angle is 21° shown in Fig. 6(a), the four corners and the short side of the rectangle holes exist enhanced electric field for RCP wave, resulting in high absorption. The enhancement of electric field originates from near-field effect between adjacent holes. For LCP wave, as shown in Fig. 6(b), it is observed that the electric field is relatively weak. The difference in electric field between LCP and RCP waves leads to strong CD. Figures 6(c) and 6(d) show the distributions of electric field when the rotation angle is 60° for RCP and LCP waves, respectively. It can be seen that there is no difference in the distribution intensity of the electric field, but the position of the field enhancement is different. Therefore, there is a weak chirality at this situation.

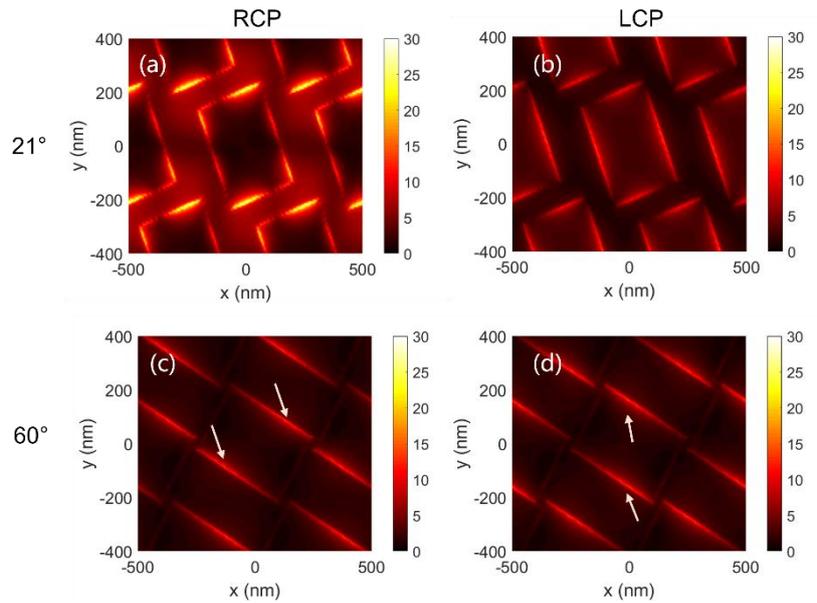

Fig. 6 The distributions of electric field in the *x-y* plane for LCP wave and RCP wave under different rotation angles: (a) $\theta$=21° for RCP wave; (b) $\theta$=21° for LCP wave; (c) $\theta$=60° for RCP wave; (d) $\theta$=60° for LCP wave.

The above results show that the orientation of the resonant unit has a great influence on CD. To explain the phenomenon mentioned above, top views of the proposed chiral structure under different rotation angles are exhibited in Fig. 7. According to the arrangement characteristics of rectangular holes, it can be found that with any one hole as the center, the line connecting the centers of six holes in its vicinity is always a regular hexagon. As we all know, chirality is closely related to the symmetry of the structure. Regular hexagon is centrosymmetric and rotationally symmetric geometry, and the included angle between two adjacent symmetrical lines is 30°, seen in Fig. 7(a). Taking Fig. 7(g) as an example to discuss the reason for weak chirality. The red dashed lines give the parallel lines of the two sides of the regular hexagon and a diagonal prior line. One can see that the red dashed lines are parallel to the long side of the rectangles and perpendicular to the short side of the rectangles. Therefore, the rectangular holes are symmetrical about the diagonal line of the regular hexagon. Due to the symmetry of the structure, the chirality is weak when the rotation angle is equal to 60°. A similar symmetry exists when the rotation angle is 0°, 30°, and 90°, respectively. However, when the rotation angle is 10°, 20°, 40°, 50°, 70°, and 80°, there is always an acute or obtuse angle between the length and width of the rectangle holes and the diagonal line of the regular hexagon, which effectively breaking the symmetry of the overall structure. Thus, the strong CD originates from symmetry breaking.

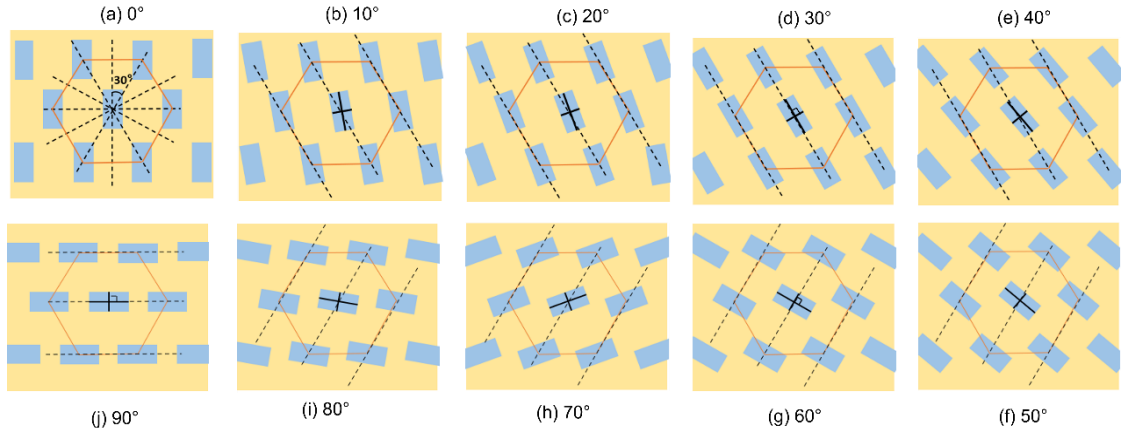

Fig. 7 Top views of the proposed chiral structure under different rotation angles: (a) 0°; (b) 10°; (c) 20°; (d)30°; (e) 40°; (f) 50°; (g) 60°; (h)70°; (i) 80°; (j) 90°.

When the connecting lines between the centers of adjacent rectangular holes is a regular hexagon, shown in Fig. 7, the CD of the proposed structure is zero with the rotation angle equal to 0 ° or an integral multiple of 30 °. As we all know, square also is a typical centrosymmetric and rotationally symmetric geometry. Thus, we speculate that when the arrangement of rectangular holes is square, the chiral response of the structure will have a similar phenomenon. Figure 8 exhibits the top view of the proposed structure when the rectangular holes are arranged square. $c$ is the period of the structure, $\beta$ is the rotation angle, $m$ and $n$ are the width and length of the rectangular holes, respectively. From Fig. 8, it can be found that the included angle of two adjacent symmetrical lines is 45°. We further speculate that when the rectangular holes arrangement is square and the rotation angle is 0 ° or an integral multiple of 45 °, CD is equal to 0.

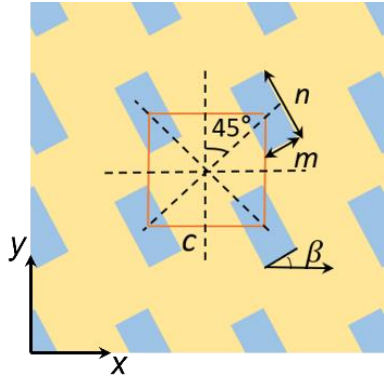

Fig. 8 Top view of the structure when rectangular holes are arranged square. $c$ is the period of the structure, $\beta$ is the rotation angle, $m$ and $n$ are the width and length of the rectangular holes, respectively.

To confirm the speculation mentioned above, we discuss the reflection and CD when the rotation angles are 20° and 45°, respectively. Here, the structural parameters of rectangular holes are the same as in Fig. 4 except for the change in arrangement. The parameters as follows: $c$=500 nm, $m$=280 nm, and $n$=400 nm. As shown in Fig. 9, there is a difference in reflection for LCP and RCP waves when the rotation angle is 20°, and the corresponding CD reaches the maximum at the wavelength of 900 nm. Although the maximum of the CD is only 0.2, the chiral response would be enhanced to suit the practical applications through optimizing the parameters appropriately. However, one can be seen that the reflection is the same for LCP and RCP waves when the rotation angle is 45°, and the CD is zero in a broadband, indicating that the chirality dose not exist under this situation.

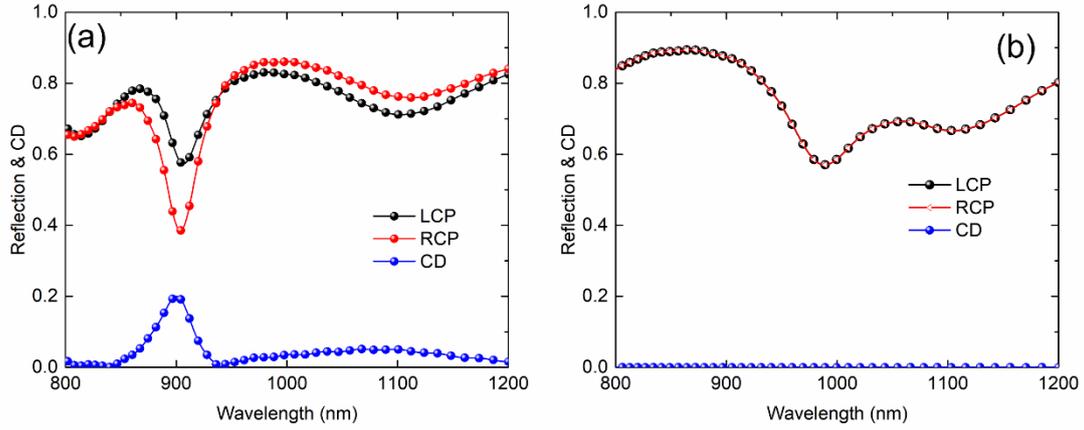

Fig. 9 The reflection and CD spectra for LCP and RCP waves when the rotation angles are (a) 20° and (b) 45°, respectively.

Next, the wavelength is fixed at 900 nm. The effect of rotation angle on reflection and CD is studied and illustrated in Fig. 10. One can see that the reflection for LCP wave incidence is greater than that for RCP wave when the rotation angle is less than 45°. However, when the rotation angle is greater than 45°, the reflection for RCP wave incidence is greater than that for LCP wave. Hance, the CD can be reversed with the rotation angle increases. In addition, it can be noted that the value of CD is 0 when the rotation angle is 0°, 45° and 90°, and CD as a function of rotation angle is symmetrical about 45°. The results are in good agreement with our speculation.

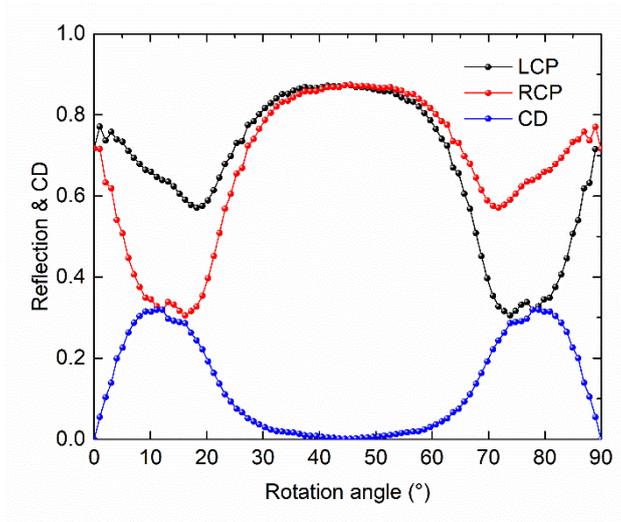

Fig. 10 The reflection as a function of rotation angle and the corresponding CD.

## 4. Conclusion

In summary, chiral metasurface based on rectangular holes has been systematically studied. The results show that strong chirality is achieved, and the maximum of CD can be up to 0.76. The physical mechanism of chirality is further explained by the electric field distributions under the illumination of LCP and RCP waves. Besides, it can be found that the period of the structure is always smaller than the wavelength at which chirality appears. In addition, it is found that the orientation of resonance unit is a key factor affecting CD. By changing the rotation angle, the overall symmetry of the structure is altered, and thus the chirality can be effectively regulated. This work will find applications in polarization-dependent photonic devices, biological sensing and analytical chemistry.

## Acknowledgements

This works is partly supported by the Training Program of the Major Research

Plan of the National Natural Science Foundation of China (92052106), the National Natural Science Foundation of China (61771385 and 52106099), the Science Foundation for Distinguished Young Scholars of Shaanxi Province (2020JC-42), the Science and Technology on Solid-State Laser Laboratory (6142404190301), the Science and Technology Research Plan of Xi'an city (GXYD14.26), the Shandong Provincial Natural Science Foundation (ZR2020LLZ004), and the Start-up Funding of Guangdong Polytechnic Normal University (2021SDKYA033).